# Transforming Software Development: Evaluating the Efficiency and Challenges of GitHub Copilot in Real-World Projects

Ruchika Pandey, Prabhat Singh, Raymond Wei, Shaila Shankar
Security Business Group, Cisco Systems Inc

## Abstract

Generative AI technologies promise to transform the product development lifecycle. This study evaluates the efficiency gains, areas for improvement, and emerging challenges of using GitHub Copilot, an AI-powered coding assistant. We identified 15 software development tasks and assessed Copilot's benefits through real-world projects on large proprietary code bases. Our findings indicate significant reductions in developer toil, with up to 50% time saved in code documentation and autocompletion, and 30-40% in repetitive coding tasks, unit test generation, debugging, and pair programming. However, Copilot struggles with complex tasks, large functions, multiple files, and proprietary contexts, particularly with C/C++ code. We project a 33-36% time reduction for coding-related tasks in a cloud-first software development lifecycle. This study aims to quantify productivity improvements, identify underperforming scenarios, examine practical benefits and challenges, investigate performance variations across programming languages, and discuss emerging issues related to code quality, security, and developer experience.

## 1 Introduction

Generative AI is rapidly transforming software development by assisting with code generation, debugging, and other tasks, thereby reshaping traditional development processes. The increasing adoption of AI tools in software engineering, particularly generative AI models like code generators and automated testing tools, necessitates a systematic study to understand their potential gains and pitfalls. The introduction of GitHub's Copilot, an AI-powered coding assistant developed by Microsoft, has prompted extensive research into its effectiveness and reliability in code generation. Previous studies have highlighted Copilot's capabilities and limitations in various programming contexts.

In this study, we aim to quantify the time savings and productivity improvements achieved by using GitHub Copilot across various software development tasks. We also seek to determine specific scenarios and types of tasks where GitHub Copilot currently struggles or underperforms. Additionally, we examine the practical benefits and challenges of integrating GitHub Copilot into real-world software engineering practices, particularly in a cloud-first development environment. Our investigation also focuses on how Copilot's performance varies across different programming languages and types of coding tasks.

Lastly, we aim to identify and discuss the emerging challenges associated with the use of generative AI tools like GitHub Copilot, including issues related to code quality, security, and developer experience.

## 2 Related Work

The introduction of GitHub's Copilot, an AI-powered coding assistant developed by OpenAI and Microsoft, has spurred extensive research into its effectiveness and reliability in code generation. Nguyen and Nadi (2022) evaluated Copilot's output for correctness and understandability using 33 queries from the LeetCode platform across four programming languages, finding that Copilot consistently produced low-complexity code, with the highest correctness in Java and the lowest in JavaScript. Sobania et al. (2021) compared Copilot's capabilities in program synthesis with genetic programming techniques, finding that Copilot's synthesized code was generally faster, more readable, and more applicable to practical scenarios.

Vaithilingam et al. (2022) conducted a user study exploring Copilot's usability in practical scenarios. While Copilot did not improve task completion time or success rate, most participants preferred it for daily programming due to its useful starting points and reduced need for online searches. However, difficulties in understanding, editing, and debugging generated code hindered its effectiveness. Arghavan Moradi Dakhel et al. (2022) assessed Copilot's capability in solving basic algorithmic problems, noting that while Copilot can generate solutions for most fundamental algorithmic tasks, some solutions are incorrect, especially for problems requiring the integration of multiple methods.

Barke et al. (2022) examined programmer interactions with Copilot, identifying two main modes: acceleration and exploration, highlighting the varied ways programmers incorporate Copilot into their workflows. Ziegler et al. (2022) analyzed the relationship between perceived productivity impacts and actual usage data, finding a strong correlation between developers' acceptance rate of Copilot's suggestions and their reports of increased productivity.

Focusing on security, Pearce et al. (2022) identified potential vulnerabilities in Copilot's code completions, finding that approximately 40% of generated code was insecure. Owura Asare et al. (2022) evaluated Copilot's effectiveness in generating correct and efficient code for fundamental algorithmic problems, comparing it to human-generated solutions and noting its standalone capabilities and relative effectiveness.

Collectively, these studies provide a multifaceted evaluation of Copilot, covering aspects such as usability, code quality, and security implications. These comprehensive assessments are essential as these tools become more integrated into software development processes, ensuring they enhance rather than compromise the quality and security of the resulting software. In this paper, we identified 15 distinct tasks within software development and assessed the benefits of GitHub Copilot adoption for each task. This novel approach enables us to delineate specific tasks where Copilot excels and those where it requires further enhancement. The study was conducted on real-world software projects, utilizing large existing codebases. Consequently, we contend that our findings hold greater relevance to practical software engineering contexts compared to studies employing synthetic coding problems. This approach ensures that the insights derived are directly applicable to enterprise software development practices, providing a robust framework for understanding the practical utility and limitations of generative AI tools like Copilot in real-world scenarios.

## 3   Methodology

The methodology of the study involved a structured evaluation with a study group of 26 engineers, covering spectrum of expertise and coding experience, from junior to senior engineers. They engaged with GitHub Copilot as part of their everyday coding tasks within a full stack cloud software development environment. GitHub Copilot was not finetuned or trained on the internal codebases used in this study. The programming languages involved included major ones like C/C++, Golang, Python, and JavaScript/PHP and the work covered a wide range of coding scenarios like new code development, debugging and refactoring, code and monitor deployment and troubleshooting, resolving customer found defects, etc. Participants were asked to use GitHub Copilot for their everyday tasks which varied in complexity and nature, thus ensuring a comprehensive assessment across different areas of the cloud software development lifecycle. Each developer maintained a detailed log of their interactions with Copilot, noting efficiency changes, challenges encountered, and the context in which the tool was used or was found lacking. The study design established a baseline for coding efficiency by comparing it to work on similar tasks but without Copilot. This comparative setup aimed to isolate the specific contributions of Copilot to coding practices, measuring its impact on productivity, and identifying areas where it either enhances or impedes the software development process. The study group spent the first week of the study getting familiar with GitHub copilot, its features and functions and different areas of work in which it could be beneficial. They also exchanged learnings on more and less effective ways to use the tool, craft prompts, feed it context, etc.  Work areas, type, and complexity of tasks as well as programming languages that would be part of the study were established for systematic data collection.

GitHub Copilot was employed across tasks as shown in Table 1.

| Software Development Task | Definition |
| --- | --- |
| New Code Development | Using prompts to have Copilot generate brand new code |
| Code Enhancements | Focused on extending functionality, improving existing features, optimizing performance, and ensuring scalability by making changes to existing functions |
| Code Auto Complete | Using Copilot functionality that predicts and completes code snippets, reducing repetitive coding tasks, minimizing syntax errors, and speeding up development time |
| Pair Programming | Using CoPilot as a pair programmer to ask questions and get help with functions and syntax in the IDE without switching to external sources like API specs, Google or Stack Overflow, etc. |
| Code Refactoring | Refactoring existing code for readability |
| Code Explanation | Providing summary, explanation, and walkthroughs of existing proprietary code, facilitating easier onboarding and knowledge transfer among team members |
| Unit Tests | Generation of unit tests for source code |
| Monitors | Generation of component tests, integration tests, and monitors |
| Code Review | Code review effort and efficiency changes due to Copilot assisted code generation |
| CI/CD | Development of Continuous Integration and Continuous Delivery pipelines that automate the testing and deployment process |
| Documentation | Creating and maintaining clear, comprehensive, and up-to-date documentation for methods and functions |
| Fixing Errors (Compilation, Linter) | Addressing and resolving syntax errors and warnings identified during compilation and by linting tools |
| Debugging | Help debugging and fixing customer issues |
| Other | Troubleshooting, text generation, format transformation, regex creating, configuration across distributed infrastructure, analysis of JSON files for simple stats, etc. |

Tasks were classified into four categories based on type and complexity.

| Task Type and Complexity | Definition |
| --- | --- |
| Repetitive | Similar tasks that need to be performed repeatedly or in multiple places |
| Boilerplate | Advanced auto-complete and code generation based on publicly available and widely understood patterns and well documented functions and APIs |
| Small function or context | Custom code but with relevant context available within a small set of open files and within the context window |
| Complex multi file | Complex custom algorithms or code that has a lot of dependencies across multiple files |

The languages used were the ones the participants use regularly for their daily tasks. This included C, C++, Golang, Python, JavaScript, JavaScript/Node.js, JavaScript/React, PHP, TypeScript. Scripting languages like bash, and config file formats like for Jenkins, Docker, YML, Ansible configs, Groovy, protobuf config, Json, xml, plain text, JMeter jmx, YAML, etc. were also used.

## 4   Results and Discussion

### 4.1   Task Type

Software development task type was the first vector used to examine potential efficiencies introduced and challenges faced with Copilot, as illustrated by Fig.1. Both varied significantly based on the type of task performed, with comment and Doxygen document generation as well as CI/CD-related tasks showing the most significant time savings at just over 50%. Deployment script generation, log analysis, and debugging were all useful in tasks related to CI/CD. Additionally, copilot helped optimize Docker code by pointing out newer and better ways to implement the same functionality. Copilot proved to be very helpful in both understanding and helping fix compiler and linter errors. Autocompletion of code, including functions, `if-else` blocks, was another area with good results and showed improvement over previous IDE autocomplete functionality. Other code generation tasks including unit test and monitor generations all had similar efficiency gains hovering around the 30-35% mark. The quality of generated codes was good for smaller and less complex tasks. Unit test generation was good for simple cases and often helped identify additional tests. One of the especially useful promising areas with Copilot was in understanding unfamiliar code and as a search and help assistant to answer questions without switching context.

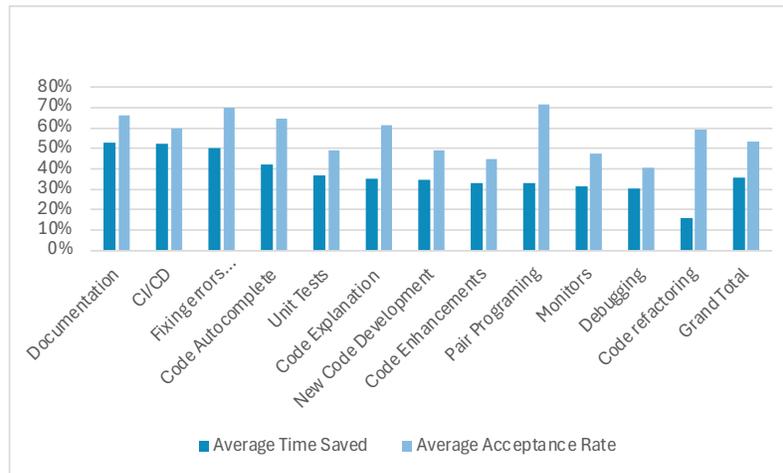

Fig. 1. Efficiency by Development Task Type

The place where Copilot code had the least acceptance and utility was with in generating code that did not follow an existing public or private pattern, especially if it involved understanding or using constructs from a large proprietary code base. For example, for a task that involved adding new display elements like buttons or toggles similar to elements that already existed in the codebase, code generation was good. Similarly, simple boilerplate actions like behavior on 'Cancel' on a file dialogue were also easily generated. But the custom functionality to be performed for other actions on display elements – for example calling APIs – needed developer intervention.

As the elements and their related actions got more complex, with a lot of different API calls and other functionality to be combined in ways that did not follow an existing pattern, the utility of Copilot code generation became a small percentage of the total work done. In other scenarios for back-end code, sometimes even when the generated code was functionally correct, it was not acceptable as it was un-optimized code. As seen in the example in Fig. 2, an expensive method was called three times instead of being called once and the value saved for future use. The code optimization use case has similar challenges. When Copilot was asked to optimize a function, it re-organized the code in such a way as to be more readable and seemingly simpler but made it less optimized. The original code conditionally called some expensive methods, while copilot's rewritten code called them unconditionally at the beginning of the function. Multiple prompts and attempts did not lead to optimized code in either the code generation or the code optimization case. For components with high performance and low latency requirements, such use cases needed careful review by domain experts to determine when the generated code could be used and when it needed to be re-written.

For unit tests, Copilot struggled in cases where objects needed to be mocked for the test. It generated comments indicating which objects and interfaces needed to be mocked but was not able to generate the relevant code. Also, different sets of tests would be generated over multiple iterations of unit test generation, some with better coverage than others. CI/CD pipeline setups required manual input for command definitions and branching conditions. Code quality could be variable, detailed prompting often helped, and a thorough review was always needed before accepting generated code. The surrounding code's quality seemed to impact the generated code's quality.  Sometimes clearing the prompt window and starting over with a better prompt was found to be useful in generating better code. For code, unit tests, as well as monitors, the quality of generated code improved with few-shot prompting – that is, if some code was first written with a pattern and data that could then be used as a reference for additional code or tests. Similarly, code generation was better if more context was provided by opening relevant files in the IDE.

```cpp
/*
 * Determines whether the data should be skipped from XYZ scanning.
 */
bool SomeClassName::skipXYZScan(const ConfigClassName::Metadata& md)) const{
//
// expensive parsing to determine if XYZ scan should be skipped
//
}

/*
 * Returns a string containing the enabled scanners for scanning.
 */
Std::string SomeClassName::getInlineScanners(const someclass::Metadata& md) {

// Custom logic

if (condition && skipXYZScan(md)) {
//do something
}

LOG(debug, SCAN_LOG_NAME "skipXYZScan = {}", __FUNCTION__, skipXYZScan(md));

if (!skipXYZScan(md)) {
//custom logic
}
//custom logic
}
```

Fig. 2.  Generated Code Example

Code quality was better if the code provided as context was of good quality, and especially if function and variable names used in the existing code were descriptive.

Copilot was helpful in accelerated understanding of unfamiliar code repos, learning unfamiliar function calls and syntax as well as publicly available API calls.

On average, the use of GitHub Copilot was observed to reduce coding time by around 35% compared to the baseline.

### 4.2   Programming Language

Another finding that emerged from the study was the wide variance in the effectiveness of GitHub copilot in different programming languages and formats, as Fig. 3 illustrates.

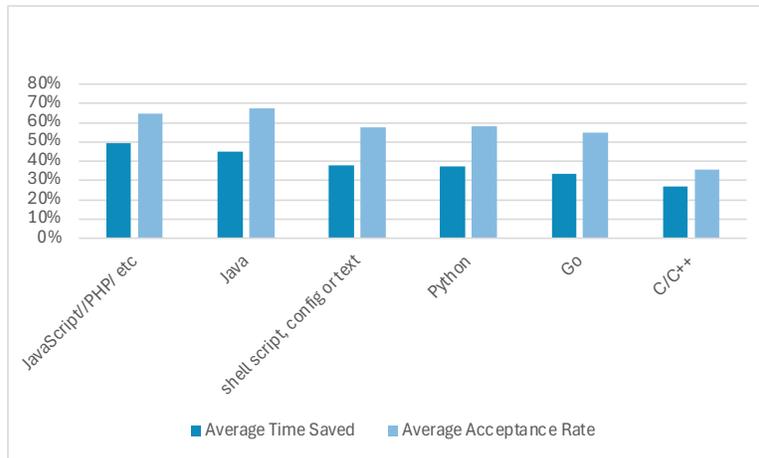

Fig. 3. Efficiency by Programming Language

Copilot performed best with JavaScript, showing time savings of close to 50%, with Java a close second at around 45%. Golang, Python and various scripting languages and configuration file updates had similar time savings of between 37% and 33%.

With C and C++, coplot often failed to generate useful code, including failing to include basic quality coding practices like null checking. Code explanation was also often found to be very basic in C and C++. The components written in C and C++ in this study are data plane components that have high performance, low latency requirements.

Copilot also gave incorrect suggestions and explanations for some Golang code that dealt with low level networking functions in Linux – it did not seem to understand the operating system specific variations in the code despite all the relevant files being open in the IDE and so arguably part of the context provided.

Fig. 4 shows the time saving for different work areas in various languages.

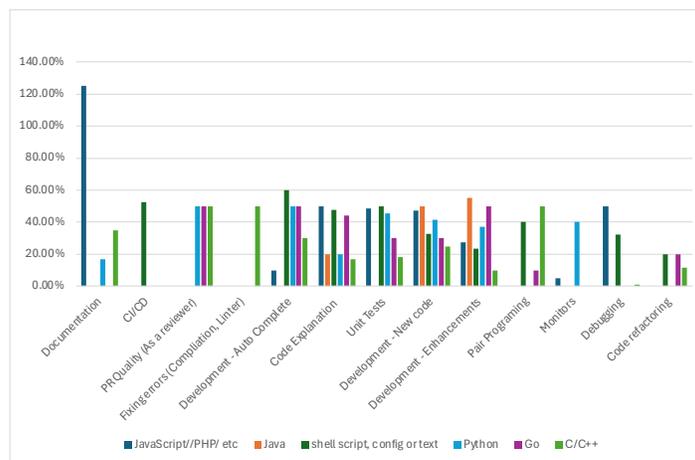

Fig. 4. Efficiency by Development Task Type

This shows that autocomplete works well in Go, Python and various scripting languages and configuration file formats. Similarly, Copilot was found to be uniformly better at new code generation in all languages, while quality was variable in updating existing code. This might potentially follow from the fact that both autocomplete, and brand-new code might have less reliance on existing local context that must be understood in updating existing code.

## 4.3   Task Complexity

Another vector along with Copilot usage was examined by type and complexity of the task. Fig. 5 shows that view.

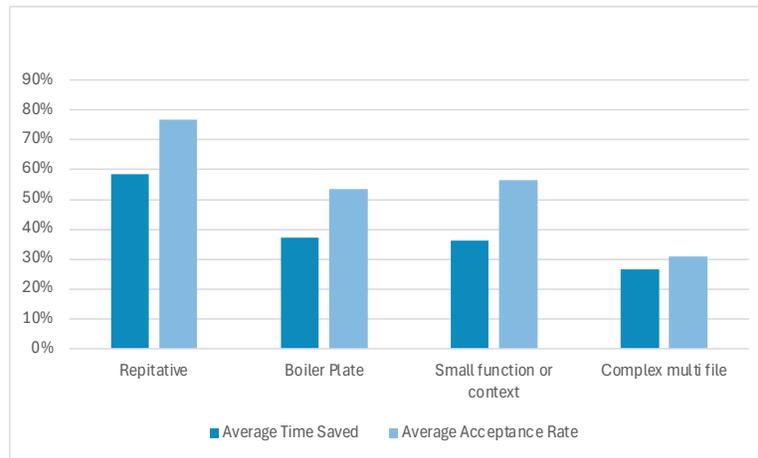

Fig. 5. Efficiency by Development Task Type and Complexity

This once again shows that Copilot does very well in saving toil related to repetitive tasks, reducing the time spent there by as much as 60%. It also does well with code generation or update when the function context is small, as well as with generating boilerplate code. The area where it does not do well is with more complex coding tasks, especially ones that involve understanding a large local context, whether that is a large function in a single file, or code spread out across many files. Complex tasks are also defined here as tasks that would need many small modifications across multiple files and functions. While Copilot chat sometimes had useful suggestions about how to approach a larger coding problem, Copilot was not found to be able to take a prompt and generate code across multiple files where modifications needed to be made to implement a complete task. In such a scenario, even if Copilot was helpful in generating code for some of the individual method or function based on detailed prompting, it was not seen as a notable efficiency gain by the study participants.

If we break tasks down further by language, we get the view in Fig. 6.

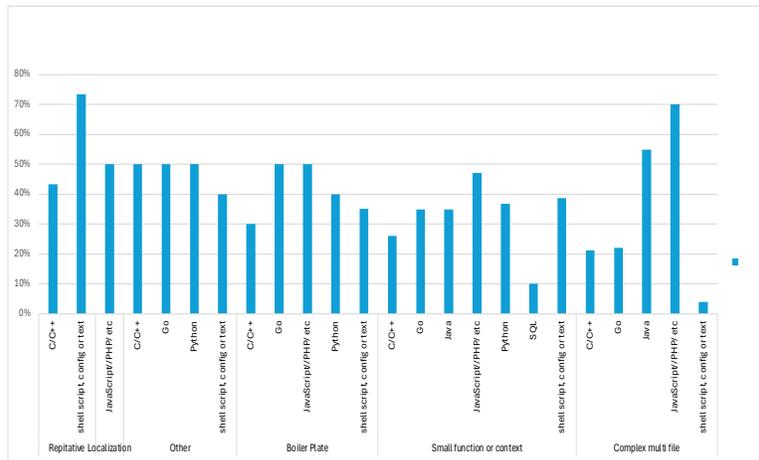

Fig. 6. Programming Language Specific Results Groped by Task Complexity

Fig. 7 shows tasks grouped by type and complexity, and average time saved for each.

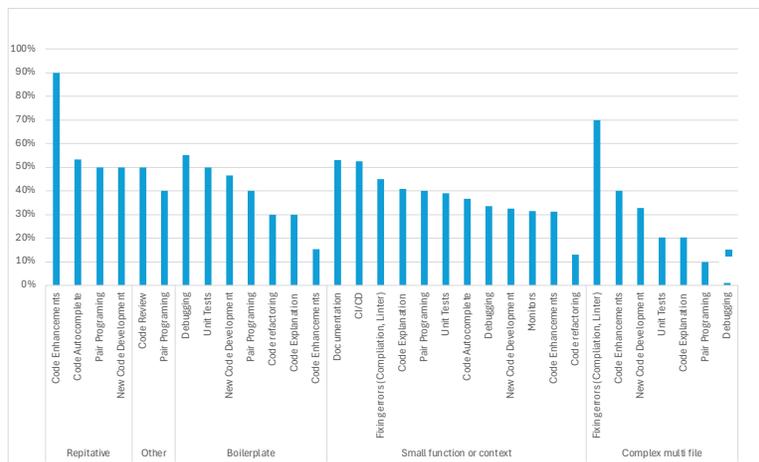

Fig. 7. Task Type Specific Results Groped by Task Complexity

This shows that the time saved using Copilot for more advanced coding tasks or tasks that require understanding a large context window or modifying code in multiple files is around 26%. For this category of tasks, Copilot is best for understanding and fixing compiler or linter errors.  This also has a bearing on the overall insights we have around time savings by using co-pilot.

## 4.4  Overall Efficiency

By modelling the approximate percentage of time spent on different work areas during feature development and the average time saved using GitHub copilot in each of these areas, we can approximate the time efficiencies that can be achieved using GitHub Copilot. This varies slightly for new components compared to feature additions and enhancements to existing components. Fig. 8 shows one possible time distribution by task type.

This includes only the areas where GitHub Copilot can be used and where it shows at least some time optimization. It does not include areas outside of that, including requirement analysis, design, dependency alignment and coordination, or test execution. All these excluded stages form a significant part of the product development lifecycle. Generative AI tools and approaches outside of GitHub Copilot would have to be explored to help speed up those stages of the lifecycle to reduce overall cycle time.

As the study progressed and participants in the test group learned what tasks copilot was good at and where it was lacking, they self-selected towards using Copilot only for the tasks where it was likely to show gains. Additionally, the model below is based on averages, and would show some amount of variance depending on the coding language, complexity of the work, etc.

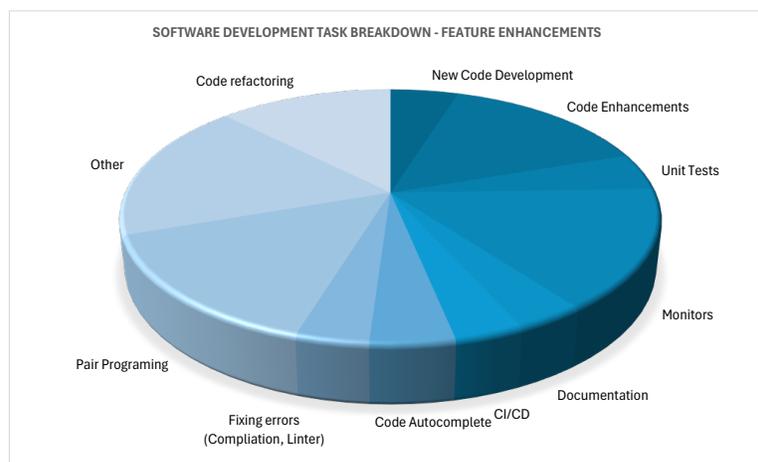

Fig. 8. Cloud Software Development Task Breakdown

Based on models created to approximate the percentage of time spent on different types of tasks in adding a feature in a SaaS product like that used in this study, an efficiency gain of between 26% and 35% could be realized by using Github Copilot.

## 5   Conclusions

### 5.1   Key Strengths

In our study on the real-world application of GitHub Copilot in cloud software development, we observed several strengths that significantly enhance developer efficiency. GitHub Copilot excels at reducing the time developers spend on repetitive and boilerplate tasks through its autocomplete functions and ability to generate relatively good boilerplate code. Its utility, however, varies by task and programming language, with Java, Golang, and Python benefiting more substantially than C and C++.

The tool also supports the generation of high-quality, consistent code comments and documentation that adhere to language conventions and style guides. Additionally, GitHub

Copilot can contribute to improved code coverage and quality by generating useful unit tests and monitors.

We noted that the quality of the generated code improves significantly when context is provided through relevant files with similar patterns, and the use of descriptive function and variable names enhances the generated code's relevance and quality. In terms of debugging compiler and linter errors, GitHub Copilot proves beneficial, especially when the problem is clearly described, leading to more targeted and useful suggestions.

GitHub Copilot also facilitates the onboarding process to new codebases by providing accurate code explanations. Similarly, it provides useful contextual help, examples and suggestions, which in turn boosts productivity by reducing the need for context switching to browser searches. Moreover, it simplifies the code review process by clearly explaining changes made to the code, thereby aiding reviewers and promoting a more standardized code quality over time.

## 5.2   Limitations and Cautions

In addition to the benefits observed with Copilot, certain important challenges and limitations were also identified.  GitHub Copilot has limited utility for advanced proprietary code, such as code that implements unique business logic, or where the relevant code is distributed over many files.  Such code often requires significant manual input. Copilot can sometimes hallucinate, but this is less of an issue since that is a problem easily caught by the compiler.  More subtle errors such as missing error checks, unoptimized code or insecure code require more careful attention, and familiarity with the programming language and the domain. Moreover, the quality of generated code can vary depending on the existing codebase. We recommend maintaining high-quality existing code, using descriptive naming, and providing detailed prompts to optimize the functionality of GitHub Copilot. Careful review and the use of other complementary tools such as static analysis and vulnerability analysis tools is currently recommended to ensure the quality and appropriateness of the generated code.

In conclusion, while GitHub Copilot presents substantial benefits in enhancing developer productivity and code quality, awareness of its limitations and careful implementation are crucial to maximizing its effectiveness in software development environments.

## 6   Future Work

The results of this study open several avenues for future research. The current study was run with Copilot without any fine-tuning on the internal codebases used in the study. The relevance of the generated code could potentially be further enhanced by performing such fine-tuning. For coding tasks, other solutions like ChatDev Chen Qian et.al and GitHub Copilot Workspace would be good candidates to explore additional efficiency gains through planning and reasoning. GitHub Copilot is only one part of the tool stack and applicable to one part of the software development lifecycle.  For future studies, additional tools and approaches can be studied, including Agentic workflows, to identify additional efficiencies that could be achieved to reduce cycle time by using Generative AI based tools.

# 7 Acknowledgements

We thank the volunteers from the Cisco Cloud Security engineering group, Nandana Bolla and Surya Allena, for their valuable contributions to this project. Their efforts in organizing the research activities and managing the data were essential to this study's success.